\documentclass[twocolumn,preprintnumbers,amsmath,amssymb]{revtex4}
\usepackage{graphicx}% Include figure files
\usepackage{dcolumn}% Align table columns on decimal point
\usepackage{bm}% bold math
\usepackage[english]{babel}
\usepackage[normalem]{ulem}
\usepackage{color}

\begin{document}
\title{Controllable supercurrent in mesoscopic superconductor-normal metal-ferromagnet crosslike Josephson structures }

\author{T.~E.~Golikova$^{a}$\email{golt2@list.ru},
M.~J.~Wolf$^{b}$, D.~Beckmann$^{b}$, G.~A.~Penzyakov$^{a}$, I.~E.~Batov$^{a}$, I.~V.~Bobkova$^{a,c,d}$, A.~M.~Bobkov$^{a}$, and V.~V.~Ryazanov$^{a,d}$}
\affiliation{$^{a}$Institute of Solid State Physics RAS, 142432
Chernogolovka, Moscow district, Russia\\$^{b}$Institute for Quantum Materials and Technologies, Karlsruhe
Institute of Technology, 76021 Karlsruhe, Germany\\$^{c}$Moscow Institute of Physics and Technology, Dolgoprudny, 141700 Russia\\$^{d}$Faculty of Physics, National Research University Higher School of Economics, Moscow, 101000 Russia}

\begin{abstract}
A nonmonotonic dependence of the  critical Josephson supercurrent on the injection current
through a normal metal/ferromagnet weak link from a single domain ferromagnetic strip has been
observed experimentally in nanofabricated planar crosslike S-N/F-S Josephson structures. 
This behavior is explained by 0-$\pi$ and $\pi$-0 transitions, which can be caused by the suppression and Zeeman splitting of the induced superconductivity due to interaction between N and F layers, and the injection of spin-polarized current into the weak link. A model considering both effects has been developed. It shows the qualitative agreement between the experimental results and the theoretical model 
in terms of spectral supercurrent-carrying density of states of S-N/F-S structure and the spin-dependent double-step nonequilibrium quasiparticle distribution.

\end{abstract} 
\maketitle
\section{Introduction}
\indent 
The interplay between spin-singlet superconductivity and ferromagnetism in mesoscopic hybrid structures leads to variety of novel effects actively investigated in the last decades. The most remarkable experiments were carried out using nonlocal technique of detection of the spin accumulation and injection in normal metals and superconductors \cite{johnson1994, Jedema1, Poli}, coherent effects of crossed Andreev reflection and elastic cotunneling \cite{deutscher2000, BeckmannCAR, Beckmann2, russo2005, Chandrasekhar}, Zeeman splitting in superconductors with adjacent ferromagnetic insulators \cite{GiazottoEuS}, charge and spin imbalance \cite{Hubler, TG2014, wolf2013, FH2012, Quay}, specific thermoelectric effects \cite{Kolenda2016, Kolenda2017, Kolenda2016_2}.

An unusual Josephson effect characterized by the inverse current-phase relation $I=-I_{c}sin\varphi$, the so called $\pi$-state of a Josephson junction, was observed in quite macroscopic trilayered SFS systems with a weak ferromagnetic interlayer \cite{SFSVR, Oboznov}. There the temperature induced 0-$\pi$ transition  was demonstrated.  
Experimentally feasible lateral superconductor (S)-normal metal (N)-ferromagnet (F)  Josephson junctions manifesting the $\pi$-state under certain conditions were studied in detail theoretically \cite{KarmPRB}. 
Implementation of such structures as submicron-scale $\pi$-phase shifters is of particular interest for superconducting and quantum electronics \cite{qbit}, which has already been demonstrated with sandwich-type SFS Josephson junctions  incorporated in superconducting logical schemes \cite{Feofanov} and in the qubit loop \cite{Scherbakova}. The future improvement is associated
with the system size reduction by means of
the fabrication
of submicron lateral hybrid structures.   
 
The driving out of equilibrium conditions makes it possible to achieve 0-$\pi$ transitions even without ferromagnets as it was predicted \cite{Wilhelm} and demonstrated \cite{Baselmans1} for controllable SNS Josephson junctions. 
The transition into $\pi$-state occurred there at a certain value of the control voltage across the N-barrier, which corresponds to a sign change in the spectrum of supercurrent-carrying states Im[$J(\epsilon)$]. The latter is characterized by the strongly damped oscillations with positive and negative parts which is cut out by the double-step-like electron distribution function or smeared by the thermal distribution function (see Fig.2 in \cite{Baselmans2}).   

It is predicted that for similar mesoscopic structures with  ferromagnets \cite{crossSFS} (or under an applied magnetic field \cite{Yip}) the spectral supercurrent is shifted due to the presence of the exchange or Zeeman field $h$ which leads to the redistribution of current-carrying states. The redistribution results in the suppression of the Josephson current in equilibrium. The recovery of the supercurrent by applying a suitable non-equilibrium distribution of quasiparticles was proposed. The recovery was predicted both for spin-independent \cite{Yip} and for spin-dependent nonequilibrium quasiparticle distribution \cite{Bobkova2012}. The proper spin-dependent quasiparticle distribution is the most efficient way to recover the supercurrent. Moreover, $0-\pi$ transitions driven by the nonequilibrium quasiparticle distribution  can be observed \cite{Yip,Bobkova2010,Bobkov2011}.  

In this article, we present the first observation of double 0-$\pi$ transition in crosslike S-N/F-S mesoscopic  \textit{Al-Cu/Fe-Al} structures.  We claim that this effect is due to spectral supercurrent modification in the presence of the effective exchange field $h$ in the hybrid weak link and controllable non-equilibrium distribution of quasiparticles. The corresponding calculations for the geometry of our structures and the parameters obtained from the measurements show qualitative agreement with the experiment.

\begin{figure}[tbp]
\centering
\includegraphics[width=0.45\textwidth]{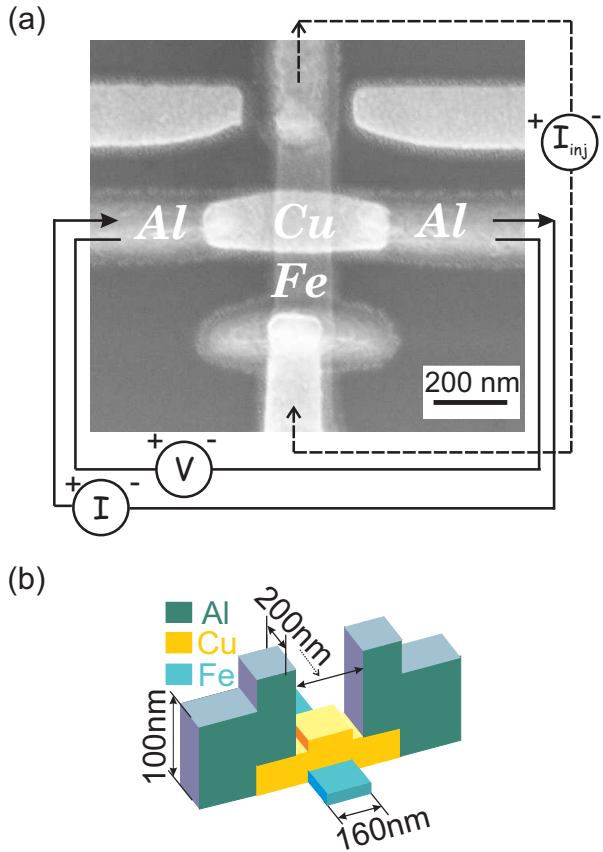}
\caption{(a) SEM image of the Al-Cu/Fe-Al junction with additional contacts to the Fe strip together with the measurement scheme. (b) The schematic sketch of the sample with geometrical dimensions.}
\label{scheme}
\end{figure}

\section{Samples and experiment}
In the investigated structures Al, Cu and Fe were used as a superconductor (S), a normal metal (N), and a ferromagnet (F) correspondingly.  
Figure \ref{scheme}(a) shows a scanning electron microscopy (SEM) image of one of our
samples, together with the measurement scheme, Fig. \ref{scheme}(b) demonstrates a schematic sketch of the structure with its geometrical dimensions. The submicron-scale crosslike S-N/F-S junctions were fabricated by means of electron beam lithography and \textit{in situ} shadow evaporation.
First, a thin (10-15 nm) iron layer was deposited onto an oxidized silicon
substrate at the first angle to form a ferromagnetic injector. Then a copper layer of  60 nm (or 30 nm) thickness was evaporated at the second angle to create a complex N/F-weak link in the intersection area of N and F layers. Finally, a thick (100 nm) layer of aluminum was deposited at the third angle to form superconducting banks of the junction and electrical contacts to the perpendicular Fe electrode.
All these technological steps were executed without breaking the vacuum, so that all FN and NS interfaces are assumed to be highly transparent. The geometrical dimensions are the same for all structures, i.e. the distance between superconducting banks is 200 nm, the width of iron strip is 160 nm, the width of copper strip is 200 nm (Fig. \ref{scheme} (b)). Since the resistance of iron film  with the specific resistance $\rho_{F}$ = 70 $\mu$$\Omega$$\times$cm and thickness $d_{F}$=10 nm is much larger than the resistance of thick copper film ($\rho_{N}$ = 4.5 $\mu$$\Omega$$\times$cm, $d_{N}$=60 nm), the main part of current flows through the N layer.

The transport measurements were performed in a shielded cryostat at temperatures down to 0.3 K. Low-pass RC filters were incorporated into DC measurement lines directly on the sample holder in order to eliminate the external noise.
The current-voltage characteristics of crosslike S-N/F-S junctions were measured by using the standard four-terminal configuration in presence of the injection current via the iron electrode as it is shown in Fig.\ref{scheme}(a).
The dimensions and thickness of the ferromagnetic
strip have provided the single domain state
with the practically uniform magnetization aligned parallel
to the the strip, as it was demonstrated
in our previous work \cite{TG2012}. 

\begin{figure}[tbp]
\centering
\includegraphics[width=0.5\textwidth]{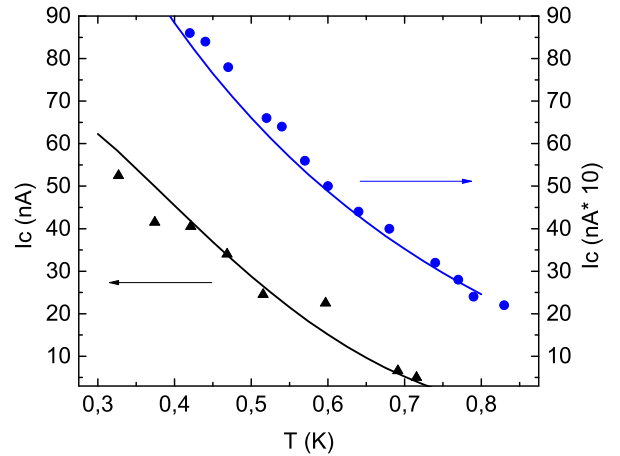}
\caption{(color online) Temperature dependencies of the critical current $I_{c}$ for the Al-Cu-Al (blue circles) and Al-Cu/Fe-Al (black triangles) Josephson junctions together with the fits (solid lines) described in the discussion.}
\label{fIcT}
\end{figure}

The Josephson supercurrent was observed only for
samples with the thickness of copper layer $d_{N}$=60 nm. In comparison with a reference S-N-S (\textit{Al-Cu-Al}) Josephson junction with the same geometrical dimensions (the only difference was the absence of a perpendicular iron strip) the critical current of hybrid S-N/F-S structures was strongly suppressed. The temperature dependencies of the critical currents for both structures are shown in Fig. \ref{fIcT}.
The suppression of the Josephson supercurrent in case of S-N/F-S structures is explained by the
proximity of the ferromagnetic layer in contact with the
copper layer in the weak link. 
Although the area of intersection of N and F layers in the cross-shaped S-N/F-S structures is much less than in the layer-on-layer weak link \cite{TG2012}, the F layer still affects considerably the transport superconducting coherence properties. The effective spin polarization is induced into the N layer due to relatively large spin diffusion length in Cu ($\lambda_{N}$=1 $\mu$m at 1 K \cite{Kimura}) in comparison with the dimensions of Cu strip in the weak link (Fig.\ref{scheme}).  A mechanism of the supercurrent suppression in such systems has already been discussed in theoretical works \cite{crossSFS,Yip} and it is related to the modification of equilibrium spectral functions by their shifting with the effective exchange energy or Zeeman field $h$ in comparison with the normal metal case. 

\begin{figure}[tbp]
\centering
\includegraphics[width=0.5\textwidth]{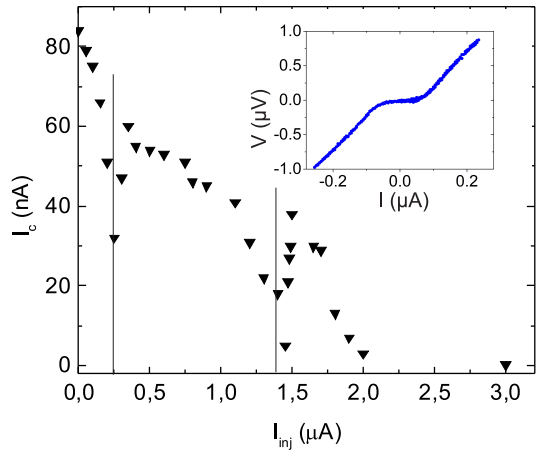}
\caption{(color online) Critical current $I_{c}$ of the crosslike Al-Cu/Fe-Al junction (A-S1) as a function of the injection current $I_{inj}$ across the junction at T=0.3 K. Black vertical lines are guides to the eye. Inset: The example of current-voltage characteristic of \textit{Al-Cu/Fu-Al} junction ($I_{inj}=0$) used for the critical current determination.}
\label{main}
\end{figure}

In the main experiments, the current-voltage (I-V) characteristics of the crosslike S-N/F-S junction were measured for different values of the injection current $I_{inj}$ across the junction (Fig. \ref{scheme}(a)). From the I-V curves we determined the critical
current $I_{c}$ for each value of the injection
current $I_{inj}$ as it is shown in Fig.\ref{main}. One can see a clear nonmonotonic
behavior of the critical current with the increasing of the injection current. This behavior does not depend on the injection current sign.     
Generally, the $I_{c}(I_{inj})$ dependence demonstrates the critical current decrease with the increasing of the injection current and two local dips at $I_{inj(1)}$=0.25 $\mu$A and $I_{inj(2)}$=1.4 $\mu$A. We suppose that the observed behavior is due to transitions between 0- and $\pi$- states at node values $I_{inj(1)}$ and $I_{inj(2)}$. As noted above, it was predicted that the different types of Josephson junctions with a complex N/F weak link can be in the $\pi$-state even in equilibrium conditions \cite{KarmPRB}, however, 
the $\pi$-state is also possible in controllable SNS systems with noneqiulibrium electron distribution due to current injection \cite{Baselmans1}.  In our work, we combine both approaches. 

\section{Discussion}

\subsection{Quasiparticle distribution in the interlayer} 

We associate the mechanism of the $\pi$-state formation with the redistribution in occupied fractions of positive and negative supercurrent-carrying density of states (SCDOS) \cite{Baselmans1, crossSFS, Yip, Bobkova2010}. For the case under consideration an effective exchange field is induced in the normal layer due to the proximity with the ferromagnet \cite{KarmPRB}. It provides more complicated SCDOS in the interlayer than for the case of a conventional S/N/S Josephson junction (see detailed discussion below). In its turn this more rich structure of SCDOS makes the Josephson current more sensitive to the quasiparticle redistribution allowing for observation of the double-transition behavior.  

\begin{figure}[tbp]
\centering
\includegraphics[width=0.5\textwidth]{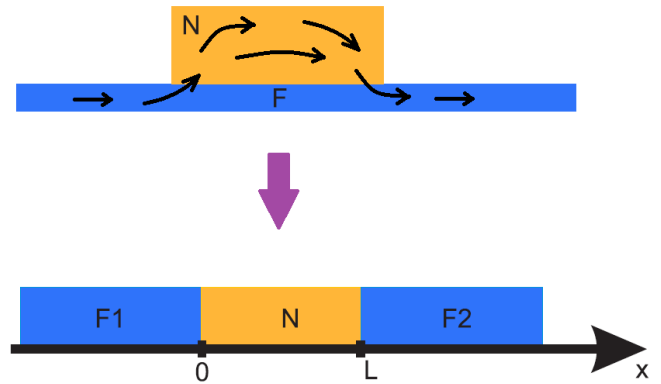}
\caption{FNF configuration for the calculation of the distribution function. The direction of the injection current flow is indicated by arrows.}
\label{FNF}
\end{figure}

At first we describe the nonequilibrium quasiparticle distribution, which is formed in the normal part of the interlayer due to the current injection from the ferromagnet in our system depicted in Fig. \ref{scheme}. For simplicity, we suppose that no transverse current enters into the superconducting banks of the junction (since the low energy process taking place at energies much less than superconducting gap of aluminum $\Delta$=180 $\mu$eV is considered, and, apart from this, the S parts are shifted away from the F one and have no intersection with it). We can consider our crosslike N/F system as F/N/F spin valve in parallel configuration (Fig. \ref{FNF}) since the main part of the injection current flows through the N layer, as it was noted above. 
      
The quasiparticle distribution in the F layers can be described in terms of the different electrochemical potentials $\mu_{\uparrow,\downarrow}$ \cite{Jedema2}, while for the N layer such  description is inappropriate because the N layer length $L$ is shorter than all the inelastic relaxation lengths and the Fermi distribution is not formed here. The spin relaxation length for Cu greatly exceeds the N layer length $L$=200 nm, as it was noted above. Therefore, the spin relaxation term can also be neglected in the kinetic equation, which can be written for the distribution function $\varphi_\sigma$ for the spin subbands separately and takes the form:
\begin{eqnarray}
\partial_x^2 \varphi_\sigma = 0.
 \label{kin_eq}
\end{eqnarray}
This equation should be supplemented by the Kuprianov-Lukichev boundary conditions at $x=0,L$:
\begin{eqnarray}
\partial_x \varphi_\sigma \Bigl |_{x=0,L} = \mp \frac{G_\sigma}{\sigma_N}\Bigl( \tanh \frac{\varepsilon-\mu_\sigma^{L,R}(x=0,L)}{2T}-\varphi_\sigma \Bigr),
 \label{bc}
\end{eqnarray}
where $\mu_\sigma^{L,R}(x=0,L)$ are electrochemical potentials for left (right) ferromagnets at the F/N interfaces, respectively and $G_\sigma$ is the conductance of F/N interfaces for the spin subband $\sigma$. The solution of Eqs.~(\ref{kin_eq})-(\ref{bc}) takes the form:
\begin{eqnarray}
\varphi_\sigma = \frac{1}{2}\Bigl( \tanh \frac{\varepsilon-\mu_\sigma^{L}}{2T}+\tanh \frac{\varepsilon-\mu_\sigma^{R}}{2T} \Bigr)+~~~~~~~~ \nonumber \\
\frac{G_\sigma}{2 \sigma_N (1+\frac{G_\sigma L} {2 \sigma_N})}\Bigl( \tanh \frac{\varepsilon-\mu_\sigma^{R}}{2T}-\tanh \frac{\varepsilon-\mu_\sigma^{L}}{2T} \Bigr)(x-\frac{L}{2}),~~~~
 \label{distribution_N}
\end{eqnarray}
where $\mu_\sigma^L$ and $\mu_\sigma^R$ are taken at the corresponding N/F interfaces $x=0,L$. The electrochemical potentials in the ferromagnets can be found from the appropriate diffusion equation and take the form \cite{Rashba2002,Jedema2}:

\begin{eqnarray}
\mu_\sigma^L (x) = A + \frac{je(x-L/2)}{\sigma_F} + \frac{\sigma C e^{x/\lambda_F}}{\sigma_F^\sigma}, \\
\mu_\sigma^R (x) = -A + \frac{je(x-L/2)}{\sigma_F} + \frac{\sigma D e^{-(x-L)/\lambda_F}}{\sigma_F^\sigma}. 
\label{mu_L}
\end{eqnarray}
where $\lambda_F$ is the spin diffusion length in the ferromagnets, $\sigma_F^\sigma$ is the ferromagnet conductivity for a given spin subband and $\sigma = \pm 1$ for the up (down) subbands, respectively. Constants $A$, $C$ and $D$ are to be found from the condition of the continuity of the electric current  at the N/F interfaces for each of the spin subbands separately. The electric currents $j_\sigma$ in the ferromagnets can be calculated as $j_\sigma^F = (\sigma_F^\sigma/e) \partial_x \mu_\sigma$, while in the normal interlayer it should be calculated according to:
\begin{eqnarray}
j_\sigma^N = \frac{\sigma_N}{4e}\int \limits_{-\infty}^\infty d \varepsilon \partial_x \varphi_\sigma (\varepsilon).
\label{j_N}
\end{eqnarray}
Then the condition $j_\sigma^F = j_\sigma^N$ at $x=0,L$ gives us constants $A$, $C$ and $D=-C$, and the electrochemical potentials $\mu_\sigma^L(x=0)$ and $\mu_\sigma^R(x=L)$ entering Eq.~(\ref{distribution_N}) take the form:
\begin{eqnarray}
\mu_\sigma^L \equiv \mu_\sigma = \frac{je\lambda_F}{(1-\kappa_\sigma)\sum \limits_\sigma \frac{\kappa_\sigma \sigma_F^\sigma}{1-\kappa_\sigma}}, \\
\mu_\sigma^R = - \mu_\sigma^L = -\mu_\sigma,
\label{mu_interfaces}
\end{eqnarray}
with
\begin{eqnarray}
\kappa_\sigma = \frac{\lambda_F G_\sigma}{2\sigma_F^\sigma(1+\frac{G_\sigma L}{2 \sigma_N})}.
\label{kappa}
\end{eqnarray}

\begin{figure}[tbp]
\centering
\includegraphics[width=0.9\linewidth]{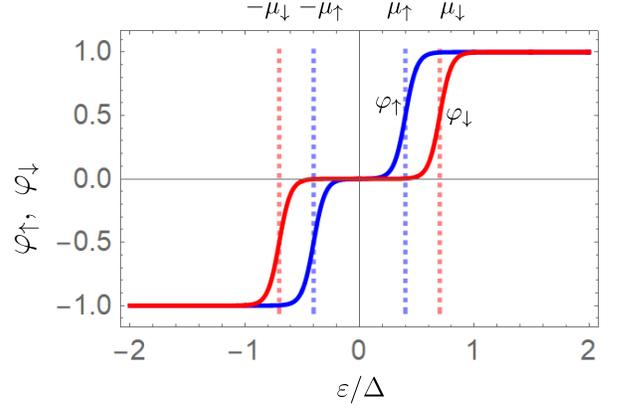}
\caption{Characteristic form of the distribution function plotted according to the first spatially constant term of Eq.~(\ref{distribution_N}).}
\label{distrib_0}
\end{figure}

Experimentally relevant parameters allow for disregarding the incline of the distribution function in the interlayer, described by the second term in Eq.~(\ref{distribution_N}). Then the distribution function can be considered as spatially constant and the characteristic example is plotted in Fig.~\ref{distrib_0} as a function of energy. It is seen that the distribution functions are different for the both spin subbands and exhibit the double-step structure, which is typical for mesoscopic regions between two leads, where the distribution function is not thermalized and is mainly determined by the distributions in the leads.

Because of the condition $\mu_\sigma^R = -\mu_\sigma^L$ the distribution function is antisymmetric with respect to $\varepsilon \to -\varepsilon$. Together with the fact that in general $\mu_\uparrow^{L,R} \neq \mu_\downarrow^{L,R}$ this leads to the conclusion that in the system under consideration two physically different modes of nonequilibrium quasiparticle distribution are nonzero. In notations of Refs.~\onlinecite{RMPBergeret,silaev2} they are $f_L = (1/2)(\varphi_\uparrow+ \varphi_\downarrow)$ and $f_{L3} = (1/2)(\varphi_\uparrow - \varphi_\downarrow)$. The first one is the energy nonequilibrium mode. In general, it is associated with the nonequilibrium distribution of quasiparticles over energy levels. For example, excess energy can be pumped into the system due to the injection current. The resulting nonequilibrium distribution can be thermalized (the quasiparticle subsystem is overheated) or not thermalized. The latter case is realized in our experiment in the particular form of the double-step structure of the distribution function. $f_{L3}$ is the so-called spin-energy mode, which can be associated with, roughly speaking, different energies pumped into spin-up and spin-down quasiparticles. If the spin-up and spin-down quasiparticle distributions are thermalized separately, it shows up as different effective temperatures for spin-up and spin-down quasiparticles. In our case of non-thermalized quasiparticle distributions it manifests itself by different widths of the double-step structure for spin-up and spin-down subsystems, see Fig.~\ref{distrib_0}. The other two possible nonequilibrium modes (the charge imbalance and spin imbalance) are not excited in the system.

 \subsection{Critical current of the junction under qusiparticle injection}
 
 Below we describe the calculations of a supercurrent in the S-N/F-S contact. We consider S as an ordinary superconductor in equilibrium with the superconducting gap $\Delta$, complex N/F weak link as a normal metal with non-equilibrium distribution and Zeeman splitting $h$ and the depairing parameter $\Gamma$. The latter parameter accounts for the leakage of the superconducting correlations into the ferromagnet and depairing there \cite{Rabinovich2019}. $d$ is the length of the N/F area (the distance between superconducting banks), $G_{SF}$ is the specific conductance of the S-N/F boundaries, $R_{NF}$ is the normal state resistance of the N/F area, $\sigma_{NF}$ is the conductivity and $D_{NF}$ is the diffusion coefficient. The calculation is performed in the framework of Usadel equations for Green's functions in the Keldysh technique. We linearize the Usadel equation in the interlayer region assuming that the S-N/F interfaces are low transparent. The assumption works quite well as it is indicated by further numerical estimates of the interface transparency.  The critical current can be expressed via the SCDOS and the distribution function as follows:
\begin{equation}
j_{c}=\frac{d}{8e R_{NF}}\int^{\infty}_{-\infty}d\varepsilon\sum_{\sigma}(\varphi_{\sigma}(\varepsilon)+\tilde{\varphi}_{\sigma}(\varepsilon))Im[J_{\varepsilon,\sigma}], 
\label{jccommon}
\end{equation}
where ${\rm Im}[J_{\varepsilon,\sigma}]$ is the SCDOS for a given spin subband with
\begin{equation}
J_{\varepsilon,\sigma}=\frac{sinh(\lambda_{\sigma}d)(G_{SF}/\sigma_{NF})^2({g_{S}^{R}}^2-1)}{\lambda_{\sigma}{\left(sinh(\lambda_{\sigma}d)+\frac{G_{SF}{g_{S}^{R}}}{\sigma_{NF}\lambda_{\sigma}}cosh(\lambda_{\sigma}d)\right)}^2}. 
\label{J}
\end{equation}
$g_{S}^{R} = -i(\varepsilon+i \delta)/\sqrt{\Delta^{2}-(\varepsilon+i\delta)^{2}}$ is the retarded Green's function and $\lambda_{\sigma}=\sqrt{-2i(\varepsilon+\sigma h+i\Gamma)/D_{NF}}$.	
 
In general, $\tilde{\varphi}_{\sigma}(\varepsilon)=-\varphi_{-\sigma}(-\varepsilon)$. In our case, due to the antisymmetry of the distribution function this leads to $\varphi_{\sigma}(\varepsilon)+\tilde{\varphi}_{\sigma}(\varepsilon)=\varphi_{\uparrow}(\varepsilon)+\varphi_{\downarrow}(\varepsilon)$, i.e. the effective quasiparticle distribution, which "occupies" the SCDOS in Eq.~(\ref{jccommon}), does not depend on spin. Therefore, only the energy nonequilibrium quasiparticle mode $f_L$ is relevant for the current situation.

\begin{figure}
\begin{minipage}[h]{\linewidth}
	\centering
	\includegraphics[width=0.9\linewidth]{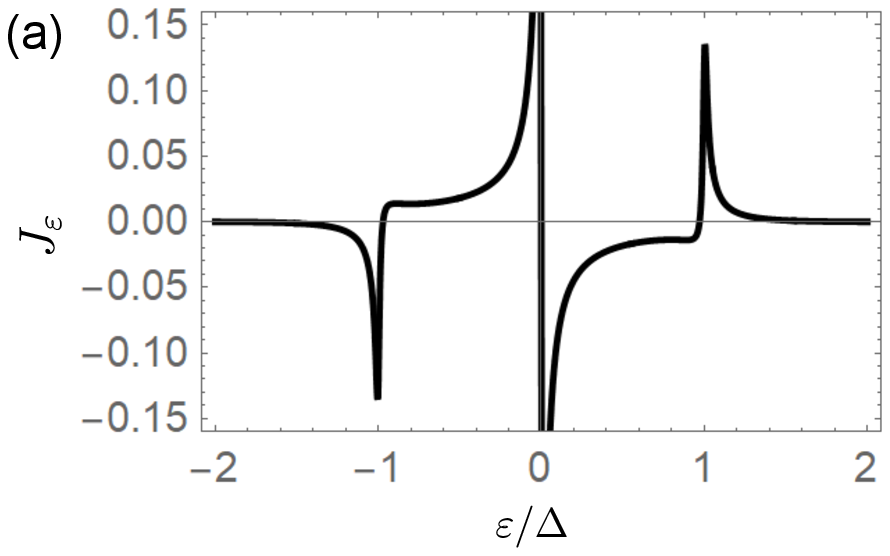}
	\includegraphics[width=0.9\linewidth]{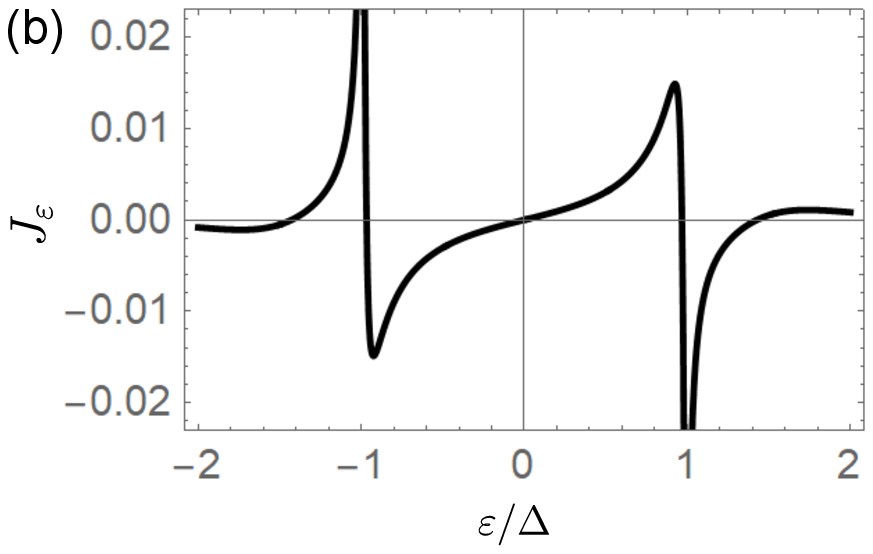}
	\caption{SCDOS $J_\varepsilon = {\rm Im}J_{\varepsilon,\uparrow}+{\rm Im}J_{\varepsilon,\downarrow}$ as a function of quasiparticle energy. (a) SCDOS for a normal interlayer with $h=\Gamma=0$; (b) SCDOS for the N/F interlayer with $h=1.41\Delta$ and $\Gamma = 0.54 \Delta$. $d = 1.76 \xi_0$ and $G_{SF} \xi_0/\sigma_N = 0.13$ for the both panels.}
	\label{SCDOS}
\end{minipage}
\end{figure}

The SCDOS for our N/F interlayer is demonstrated in Fig.~\ref{SCDOS}(b). The plot corresponds to the parameters $\tilde G = G_{SF}\xi_0/\sigma_{NF} = 0.13$, $d=1.76\xi_0$, $h=1.41 \Delta$ and $\Gamma = 0.54 \Delta$, where $\xi_{0}=\sqrt{D_{NF}/\Delta}\approx 190$ nm. These parameters are found by fitting the experimental data presented in Fig.~\ref{fIcT}.  $d$ is an effective length of the normal interlayer, it does not exactly coincides with the actual length of the Cu region because the Cu regions underneath the Al leads and Cu regions between the Al leads and the Fe strip are partially proximitized by the Al superconductivity.

The SCDOS for the N/F interlayer should be compared to the SCDOS for a usual N interlayer corresponding to $h=0$, $\Gamma=0$ [Fig.~\ref{SCDOS}(a)]. It is seen that the SCDOS is a sign changing function. This is the reason providing the possibility for $0$-$\pi$ transitions driven by an external parameter controlling the quasiparticle distribution. For the case of only one zero-crossing point at $\varepsilon>0$, as it takes place for the SCDOS in the N interlayer [Fig.~\ref{SCDOS}(a)], no more than one  $0$-$\pi$ transition driven by injection, which results in the double-step or overheated quasiparticle distribution, is possible. This situation has been realized experimentally \cite{Baselmans1}. On the contrary, the SCDOS for our system manifests two zero-crossing points at $\varepsilon>0$ due to the Zeeman splitting. This results in the possibility of two $0$-$\pi$ transitions driven by injection. 

\begin{figure}
\begin{minipage}[h]{\linewidth}
	\centering
	\includegraphics[width=0.9\linewidth]{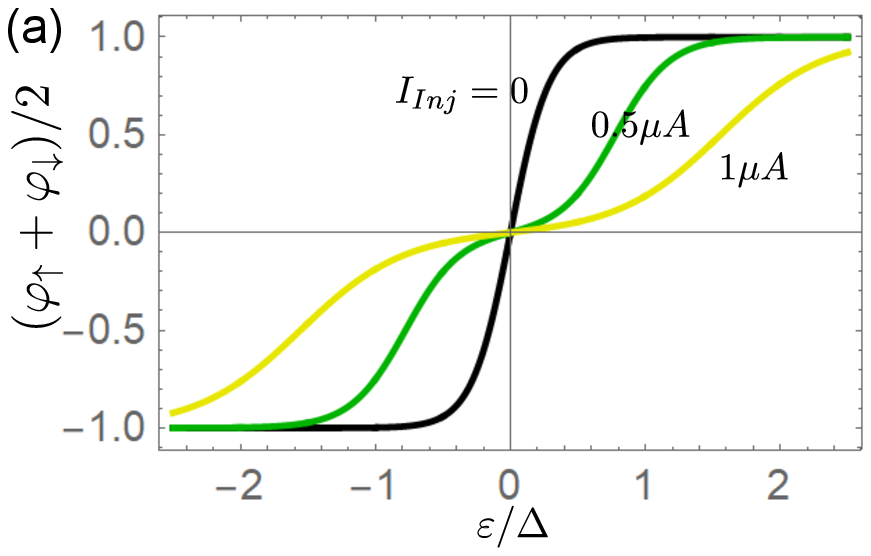}
	\includegraphics[width=0.9\linewidth]{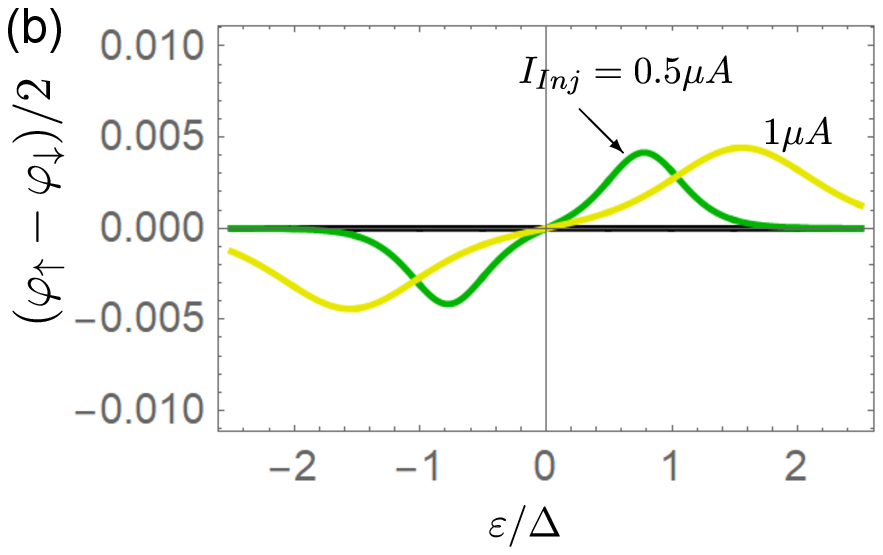}
	\caption{Quasiparticle distribution: nonequilibrium modes (a)$f_L = (\varphi_\uparrow+\varphi_\downarrow)/2$ and (b) $f_{L,3} = (\varphi_\uparrow-\varphi_\downarrow)/2$ calculated for the current experiment. $(\sigma_F^\uparrow-\sigma_F^\downarrow)/\sigma_F = 0.45$, $G_{NF} = G_\uparrow + G_\downarrow = 0.15 G_{SF}$,  $(G_\uparrow - G_\downarrow)/G_{NF} = 0.2$. $T= 0.3K$, $\beta = 0.5K/(\mu A)^2$.}
	\label{distrib_1}
\end{minipage}
\end{figure}

The evolution of the distribution modes $f_L$ and $f_{L3}$ upon increasing of the injection current is presented in Figs.~\ref{distrib_1}(a,b), respectively. As it was discussed above for the problem under consideration, the critical current is only determined by $f_L$. At zero temperature it manifests  a four-step structure as a function of energy. The steps occur at $\varepsilon = \pm \mu_\uparrow, \pm \mu_\downarrow$. According to Eq.~(\ref{mu_L}), the electrochemical potentials are proportional to the injection current $j$. Therefore, $\mu_{\uparrow,\downarrow} = \alpha_{\uparrow,\downarrow} J_{inj}$, where taking the parameters of our F/N/F structure $(\sigma_F^\uparrow-\sigma_F^\downarrow)/\sigma_F = 0.45$ \cite{Soulen,Stockmaier}, $G_{NF} = G_\uparrow + G_\downarrow = 0.15 G_{SF}$,  $(G_\uparrow - G_\downarrow)/G_{NF} = 0.2$, $\lambda_{F}$=8.5 nm \cite{Fe} we obtain $\alpha_\uparrow = 1.549\Delta/\mu A = 279eV/A$ and $\alpha_\downarrow = 1.563\Delta/\mu A =  281eV/ A$. It is seen that $\mu_\uparrow$ and $\mu_\downarrow$ are very close for the particular parameters of the structure. As a result, the spin-energy mode $f_{L3}$ is rather small here, as it is demonstrated in Fig.~\ref{distrib_1}(b) and the four-step structure is very close to the double-step one even at $T=0$. Due to nonzero temperature this step structure is smeared. The temperature smearing is further increased because of the Joule heating by the injection current, which is modelled by $T \to T+\beta J_{inj}^2$ with $\beta = 0.5K/(\mu A)^2$.

As it is seen from Fig.~\ref{distrib_1}(a), upon increasing the injection current the quasiparticles are redistributed to higher energies leaving the low-energy states. This leads to gradual turning off the low-energy parts of the SCDOS. Therefore, the main part of the SCDOS contributing to the critical current changes sign upon increasing the injection current and the maximal number of $0$-$\pi$-transitions is given by the number of zero-crossings in the SCDOS, as it was already mentioned above.

\begin{figure}
\begin{minipage}[h]{\linewidth}
	\centering
	\includegraphics[width=0.9\linewidth]{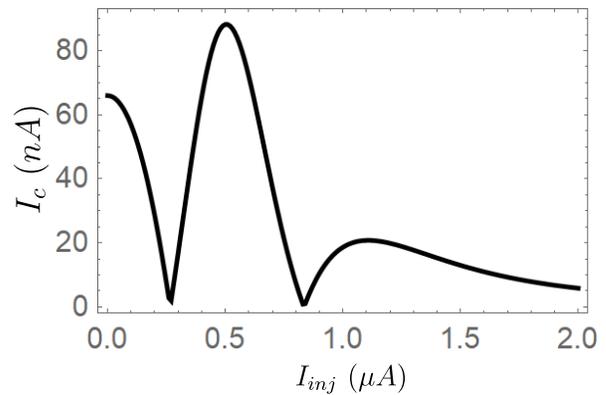}
		\caption{Critical current of the S-N/F-S Josephson junction calculated as a function of the injection current. The parameters of the junction correspond to SCDOS shown in Fig.~\ref{SCDOS}(b) and distribution function presented in Fig.~\ref{distrib_1}.}
	\label{crit_current}
\end{minipage}
\end{figure}

The critical current calculated in the framework of our theoretical model as a function of the injection current is presented in Fig.~\ref{crit_current}. It manifests two $0$-$\pi$-transitions in qualitative agreement with the experimental results. However, the quantitative agreement is lacking because of the complicated geometry of the experimental structure. To obtain better approximation one should take into account that the real junction is not a simple S-N/F-S junction but contains regions of proximity superconductivity underneath the superconducting leads, which are very sensitive to the nonequilibrium quasiparticle distribution in the interlayer.

\section{Conclusion}

To conclude, we have demonstrated experimentally the double 0-$\pi$ transition in the crosslike S-N/F-S Josephson junctions driven in non-equilibrium by applying an injection current across the complex weak link. We ascribe this effect to the appearance of two zero-crossing points in the supercurrent-carrying density of states (SCDOS) caused by the Zeeman splitting of the superconducting correlations in the N/F interlayer.  The model taking into account SCDOS of a S-N/F-S structure and spin injection into the N layer is developed to describe the observed effect. It has been concluded that the non-equilibrium in our case is not thermal, mainly described by the energy mode with a small amount of the spin-energy nonequilibrium mode without a contamination of charge and spin imbalance.
Thus, we have shown that the Zeeman splitting due to N/F proximity has provided the origin of two nodes of the alternating SCDOS, which is practically impossible to realize in simple SNS structures without the F-sublayers used in the pioneering works \cite{Baselmans1,Baselmans2}.
In our particular case, the spin-energy mode turned out to be
negligible. To demonstrate the spin-energy mode manifestations, it is
necessary that the distribution function has a spin splitting (i.e. the
difference between $\mu_{\uparrow}$ and $\mu_{\downarrow}$ curves in Fig.\ref{distrib_0}) significantly exceeding its temperature smearing.
Our consideration can be applied for a general situation with the step-like distribution function by fabricating an appropriate structures and further temperature decrease, which, in particular, can be used for determination of the effective exchange field of the hybrid S-N-F structures with different transparency of interfaces.    

The work was partially supported by grants of Russian
Academy of Sciencies and Russian Foundation for Basic
Research N 20-02-00864, 19-02-00452, 18-52-45011 and
18-02-00318, as well as by the Basic Research program of
the Higher School of Economics.

\end{document}